# Iron pnictides: Single crystal growth and effect of doping on structural, transport and magnetic properties

G. L. Sun[1], D. L. Sun[1], M. Konuma[1], P. Popovich[1], A. Boris[1], J. B. Peng[1],
K.-Y. Choi[2,3], P. Lemmens[3] and C. T. Lin[1*]

[1] *Max-Planck-Institut für Festkörperforschung, Heisenbergstr. 1, D-70569 Stuttgart, Germany*
[2] *Department of Physics, Chung-Ang University, 221 Huksuk-Dong, Dongjak-Gu, Seoul 156-756, Republic of Korea*
[3] *Institute for Condensed Matter Physics, TU Braunschweig, Mendelssohnstr. 3, D-38106 Braunschweig, Germany*

**Abstract**

We demonstrate the preparation of large, free standing iron pnictide single crystals with a size up to 20 x 10 x 1 $mm^3$ using solvents in zirconia crucibles under argon atmosphere. Transport and magnetic properties are investigated to study the effect of potassium doping on the structural and superconducting property of the compounds. The spin density wave (SDW) anomaly at $T_s$~138 K in $BaFe_2As_2$ single crystals from self-flux shifts to $T_s$~85 K due to Sn solvent growth. We show direct evidence for an incorporation of Sn on the Fe site. The electrical resistivity data show a sharp superconducting transition temperature $T_c$~38.5 K for the single crystal of $Ba_{0.68}K_{0.32}Fe_2As_2$. A nearly 100% shielding fraction and bulk nature of the superconductivity for the single crystal were confirmed by magnetic susceptibility data. A sharp transition $T_c$~25 K occurred for the single crystal of $Sr_{0.85}K_{0.15}Fe_2As_2$. There is direct evidence for a coexistence of the SDW and superconductivity in the low doping regime of $Sr_{1-x}K_xFe_2As_2$ single crystals. Structural implications of the doping effects as well as the coexistence of the two order parameters are discussed.

*Corresponding author: Tel: 49-711-6891458, Fax: 49-711-6891093,
Email: ct.lin@fkf.mpg.de

## 1. Introduction

The recent discovery of iron-based superconductors [1-3] has renewed the interest in the field of superconductivity. It has been reported that the F-doped $LaFeAsO_{1-x}F_x$ shows a superconducting $T_c$=26 K [1] and SmFeAsO reaches transition temperatures as high as 55 K under pressure [4]. Moreoever, a simpler class of materials based on the $BaFe_2As_2$ (112) parent compound that does not contain oxygen also shows superconductivity with $T_c$~38 K when potassium is doped [5]. The iron arsenides are considered as the second important class of high-$T_c$ superconductors since the discovery of the cuprates about two decades ago.

To study superconductivity, electrical and magnetic anisotropic behavior and inelastic neutron diffraction experiments of these novel compounds, single crystals are desperately required. However, the crystal growth of these materials is hampered by the toxicity of arsenides as well as the high vapor pressure of arsen and potassium. High pressure techniques were therefore employed to synthesize oxygen-containing arsenides. Polycrystalline $NdFeAsO_{0.9}F_{0.1}$ pellets containing single grains as large as 300 μm were synthesized under the high pressure of 3.3 GPa [6]. Free standing crystals of $SmFeAsO_{1-x}F_y$ with a size of up to 120×100 $\mu m^2$ were grown from salt solutions under high pressure of 30 kbar [7]. The growth of oxygen-free iron arsenide single crystals is reported using Sn or self flux method [8, 9] and a reasonable size was achieved. The single crystals of $Ba_{1-x}K_xFe_2As_2$ (x=0 and 0.45) with a size of ~3 $mm^2$ were obtained using Sn flux in quartz container [9]. A self-flux method was also developed for the growth of $BaFe_2As_2$ with size up to 5 mm [8]. Due to the high vapor pressure of the compound and corrosion of quartz containers at high growing temperatures they tend to explode hampering the reproducibility of crystal growth. Furthermore, droplets of the residual flux adhere to the crystal surface, leading to a contamination or damage of the grown crystals. Therefore, the growth of single crystals, particularly the superconducting doped variants with large size, remains a challenge.

The parent compounds of $AFe_2As_2$ (A=Ba, Sr) are intermetallic and non-superconducting. One important finding is the existence of a spin density wave (SDW) anomaly at $T_s$~140 K for $BaFe_2As_2$ and ~200 K for $SrFe_2As_2$, respectively [10, 11]. This SDW is linked to abrupt changes in the electrical resistivity and magnetic susceptibility as well as a structural transition from a high temperature tetragonal to a low temperature orthorhombic phase [10-13]. By the change of electron count in the FeAs layers the compounds become superconducting, which is believed to be associated with the suppression of SDW anomaly [12, 14, 15]. However the $T_s$ to the SDW does not manifest itself so clearly in $BaFe_2As_2$ crystals grown by Sn solvent, which is believed to be included into crystals and leads to a reduction of the phase transition temperature to 85 K [9]. This effect of Sn on $T_s$ and the local structure is not understood so far. It is possible that Sn replaces Fe in the layers. Sn could form bonds with other atoms or be included neutral in the crystal structure. Here we give strong evidence for an exchange of Fe by Sn with a bonding energy estimated by atomic phonon spectroscopy. By the doping of potassium, the transition near 85 K disappears and instead superconductivity occurs with a transition temperature of ~38.5 K probed by resistivity and magnetization measurements.

In this work we present centimeter-sized crystals obtained using either Sn or self-flux method and by a specially designed apparatus. The pure and doped single crystals were characterized with respect to their anisotropy and with respect to crystallographic, magnetic and transport properties.

## 2. Experimental

As source materials pure elements of Ba/Sr, K, Fe, As and Sn were used in a mol ratio of $[(Ba/Sr)_{1-x}K_xFe_2As_2]$:Sn=1:45-50 or in the mol ratio of $1((Ba/Sr)_{1-x}K_x)$:5Fe:5As for the self flux. Usually 90 g of the mixtures with Sn or 15 g for self-flux were loaded in a zirconia crucible. A crucible with lid was used to minimize the evaporation loss of K as well as As during growth. The crucible was sealed in a quartz ampoule filled with Ar and loaded into a box furnace. It is noticed that free-standing crystals could be obtained after decanting if holes are drilled into the crucible lids to allow the residual flux flowing out of the crucible. The details of a specially designed furnace used for decanting flux containing poisonous arsenic compounds at high temperature are shown in Fig. 1.

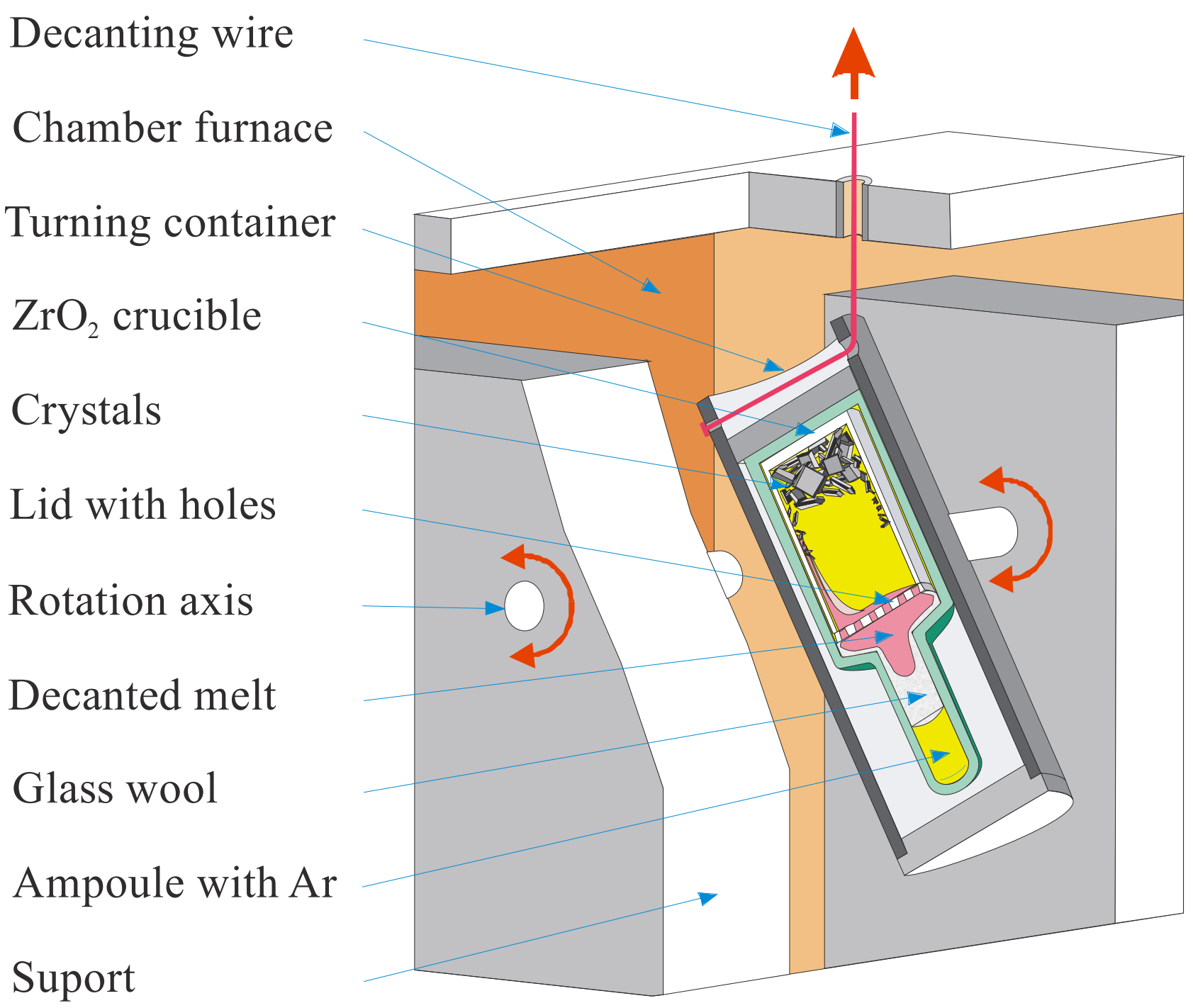

Fig. 1: A schematic drawing of the apparatus used to grow single crystals of $A_{1-x}K_xFe_2As_2$ (A=Ba, Sr). Free standing crystals are observed to dominantly grow on the bottom of the crucible after the flux is decanted.

The crystals were identified by X-ray powder diffraction (XRD) using Cu Kα radiation with a scanning rate of 0.02 °/min. The chemical composition was determined by energy dispersive X-ray spectrometry (EDX) as well as inductively coupled plasma spectroscopy (ICP). The error range for both is within 1 at%. The crystal morphology was studied by a scanning electron microscopy (SEM). The temperature dependence of electric resistivity for the pure and K-doped single crystals was investigated using a standard four-probe technique measured from 4 to 300 K. The ac magnetic susceptibility was performed by a superconducting quantum interference device magnetometer (SQUID) in different fields. High resolution electron spectroscopy for chemical analysis (ESCA) was employed to determine the binding energy of which Sn is positively charged in the crystal. Raman scattering experiments have been performed in quasi-back scattering and (xx) polarization where the incident and scattered light are polarized parallel to the *a* axis.

## 3. Results and discussion

### *3.1 Thermal behavior*

The melting behavior of pure and doped $Ba_{1-x}K_xFe_2As_2$ single crystals was investigated by thermogravimetric and differential thermal analysis (TG-DTA) with NETZSCH STA449C equipment. Tiny crystals were loaded in a lid-covered $Al_2O_3$ crucible and then heated at 5 °C/min in Ar with the flowing rate of 45 ml/min. A maximum heating temperature of 1100 °C was used for doped $Ba_{0.72}K_{0.28}Fe_2As_2$ and 1300 °C for pure $BaFe_2As_2$, respectively. The results are shown in Fig. 2. The little endothermic peaks at $T_1$= 232 °C on the DTA curves correspond to the Sn melting point temperature, indicating the existence of residual flux in both pure and doped specimens. The endothermic peaks are observed at $T_2$=917 °C for the K-doped $Ba_{0.72}K_{0.28}Fe_2As_2$ and $T_4$=1030 °C for the pure $BaFe_2As_2$, corresponding to the incongruent melting of both compounds. An exothermic effect was recorded on the DTA curves with the onset at $T_3$=988 and $T_5$=1093 °C for the doped and pure smaples, respectively. Correspondingly, on the TG curve the weight loss is observed starting from $T_3$ or $T_5$, which indicates a continuous decomposition with a loss of K and As. The onset decomposition temperature is about 100 °C lower for the doped than for the pure sample. Our TG-DTA data indicate an incongruent melt for both the doped and undoped compounds. Therefore the flux method is suggested for crystal growth.

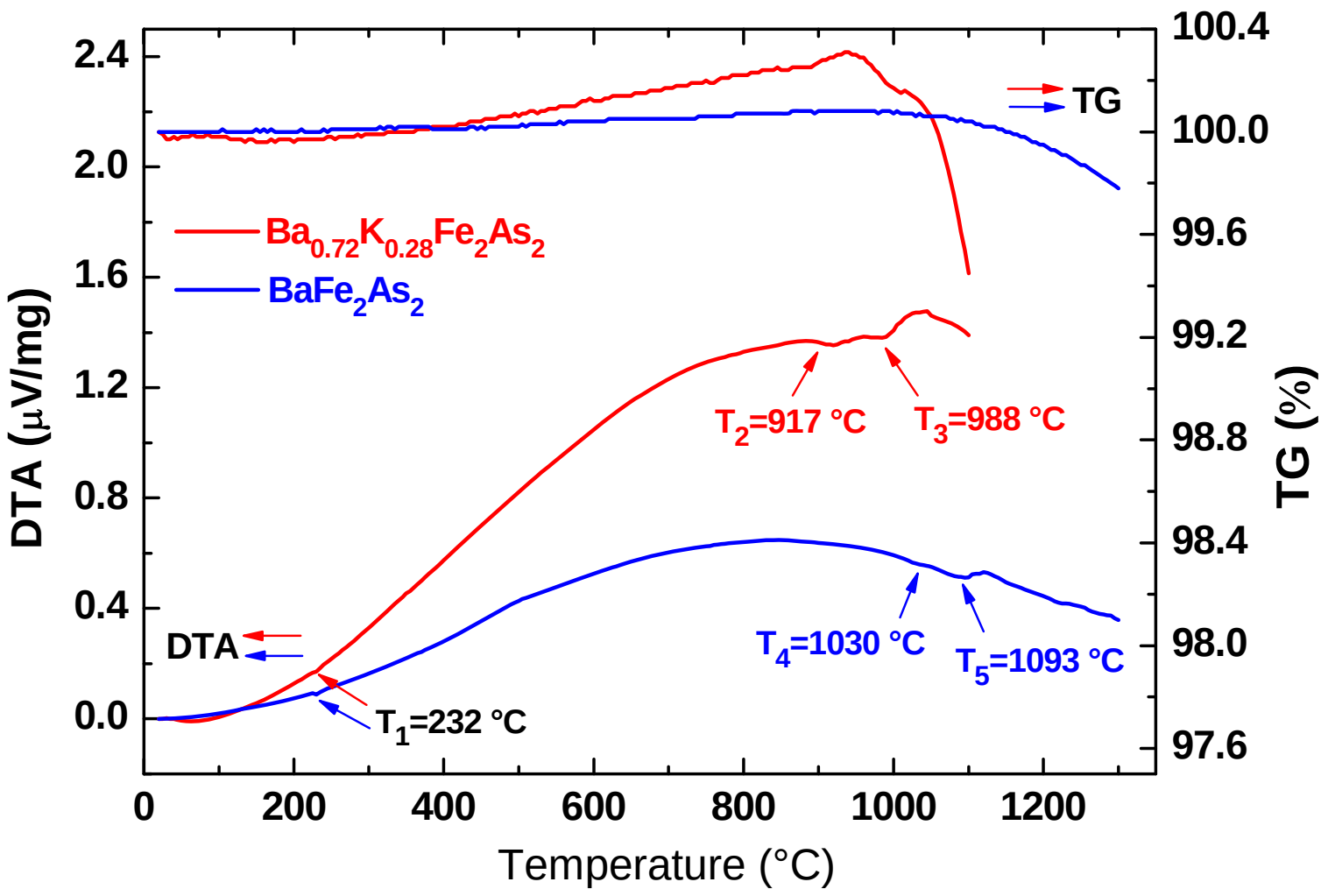

Fig. 2: Analysis of the thermal behavior for single crystals of $BaFe_2As_2$ and $Ba_{0.72}K_{0.28}Fe_2As_2$, respectively. The samples were heated at a rate of 5 °C/min with an Ar flowing rate of 45 ml/min.

*3.2 Crystal growth*

Large and high-quality crystals could be obtained in a homogeneous melt by slow cooling. To achieve a homogeneous melt the maximum heating temperatures of 980 and 850 °C were applied for soaking the pure and doped compounds with Sn flux, respectively. For the self-flux growth the heating temperatures of 1190 and 1080 °C were used for pure and doped crystals, respectively. The soaking temperatures were then maintained for 2-4 hours. Subsequently, the temperatures were decreased with a low cooling rate of ~3 °C/h down to ~550 °C for the Sn-flux and ~890 °C for the self-flux. As-grown crystals were then decanted from the flux. The free standing crystals were observed to grow dominantly at the bottom of crucible. It is noticed that to obtain crystals entirely free from flux the decanting temperature had to be maintained for 2 h with a subsequent cooling to room temperature. This allows the residual flux to flow out completely and to leave free standing crystals inside of the crucible. It is emphasized that the decanting device is designed with a movable nickel wire for tilting the crucible to drop the residual flux at high decanting temperature on the top of outside the furnace. This prevents any poisoning from As in case of a crack of the quartz tube. As-grown single crystals are shown in Fig. 3(a). Fig. 3(b) shows typical crystals with a partly "opened" surface layer, which was obtained by a rapid drop of the high decanting temperature and sudden cooling to room temperature. Besides, the layered structure with weak bonding along the *c* direction may cause crystal layers readily cracked by any thermal shock. Details of the crystal habit are described in section 3.4.

The dimensions of the crystals are as large as 20 x 10 x 1 $mm^3$, as shown in fig 3(c). The average aspect ratio of crystal dimensions is estimated and given in Table 1. Potassium doped crystals show the highest aspect ratio of up to 100:1, while the undoped ones show smaller aspect ratios and are thicker along the *c* axis direction. The compositions of the crystals were determined to be (Ba/Sr+K):Fe:As=1:2:2, within an error range of ~1 at%. It is estimated that the content of K in the crystals is ~60 mol% for $Ba_{1-x}K_xFe_2As_2$ and ~30 mol% for $Ba_{1-x}K_xFe_2As_2$, compared to the initial doping contents, respectively.

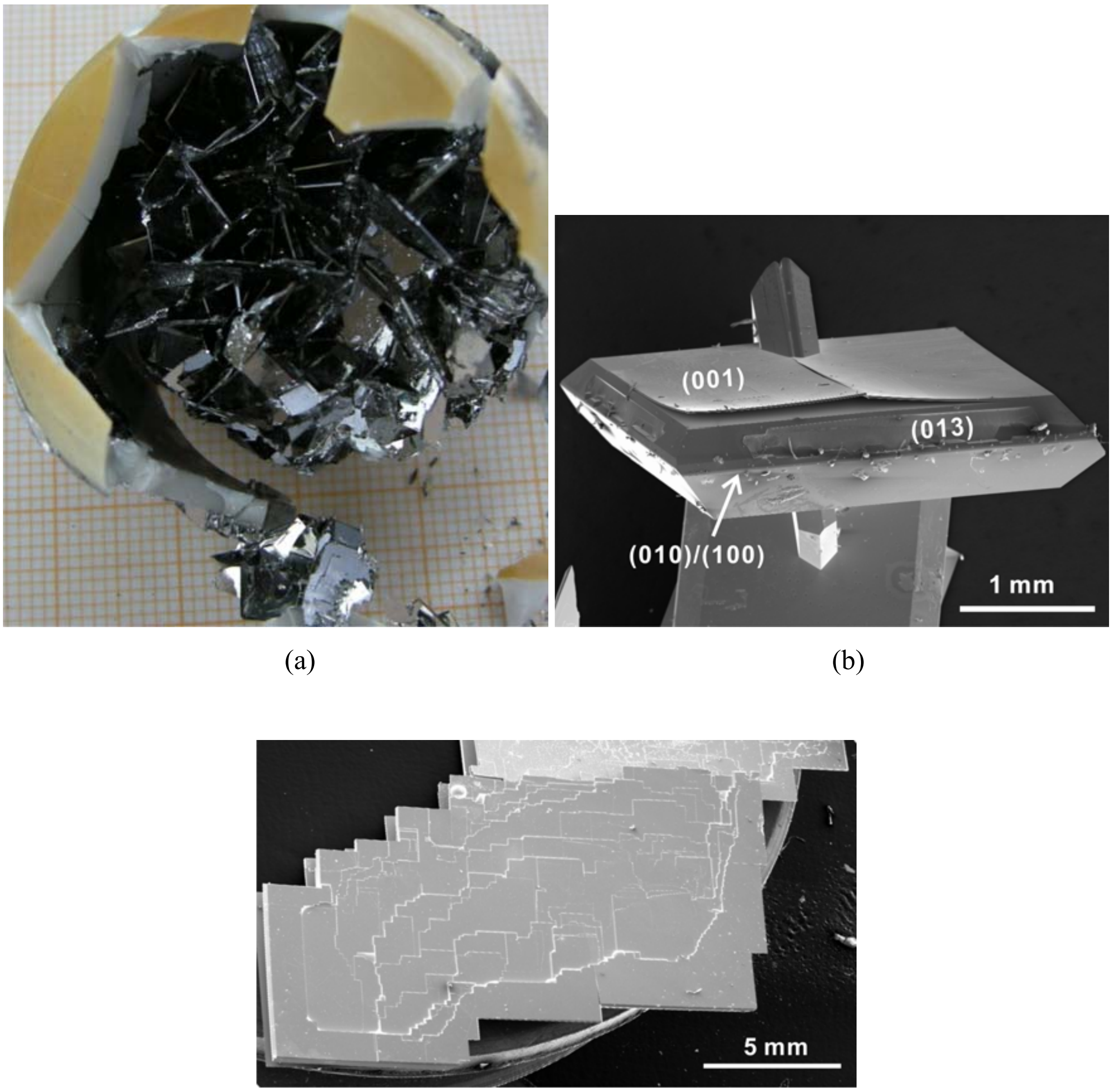

(a) (b)

(c)

Fig. 3: (a) Harvested single crystals of $BaFe_2As_2$ obtained from Sn-flux. (b) typical separated single crystals showing a partly exposed top layer (*001*) and well developed crystallographic faces. (c) A large single crystal separated from the bottom of the crucible.

*3.3 Structural analysis*

The structure and phase purity of the as-grown crystals were investigated by XRD. The XRD diffraction patterns were obtained using ground powders of as-grown single crystals. Results are shown in Figs. 4(a) and (b) for the pure and doped $Ba_{1-x}K_xFe_2As_2$ and $Sr_{1-x}K_xFe_2As_2$, respectively.

The lattice parameters derived from the powder XRD data are given in Table 1. The crystals belong to the tetragonal structure with space group *I4/mmm* at room temperature. These results are in agreement with the reported data [16-18]. Comparing the pure and doped $A_{1-x}K_xFe_2As_2$ (A=Ba, Sr), the *c* lattice parameter expands while *a* shrinks slightly with doping of potassium. The effect of potassium doping on the *c* lattice parameter is also exhibited by the (*00l*) XRD patterns taken from the crystal plane, as shown in Figs. 4(c) and (d), respectively. All the (*00l*) patterns for the doped samples shift to lower 2θ angles, compared to the undoped ones. This is indicative of an expansion of the *c* lattice parameters with potassium doping, because the ionic ratio of $K^+$ (1.51 Å) is larger than that of $Ba^{2+}$ (1.42 Å) and $Sr^{2+}$ (1.26 Å). The *c* axis lattice parameter for $BaFe_2As_2$ is larger than that of $SrFe_2As_2$ since $R_{Ba^{2+}} > R_{Sr^{2+}}$.

Table 1: The dimensions, aspect ratio and lattice parameters of $A_{1-x}K_xFe_2As_2$ (A=Ba, Sr) crystals grown from Sn-flux

| Single crystal | Typical dimensions ($mm^3$) | Aspect ratio | Lattice constant (Å) | |
|---|---|---|---|---|
| | | | *a* | *c* |
| $BaFe_2As_2$ | 10×8×1 | 6:1 | 3.9241 | 13.1536 |
| $Ba_{0.72}K_{0.28}Fe_2As_2$ | 15×10×0.3 | 25:1 | 3.9186 | 13.3227 |
| $SrFe_2As_2$ | 8×6×1 | 9:1 | 3.9386 | 12.4559 |
| $Sr_{0.85}K_{0.15}Fe_2As_2$ | 4×2×0.5 | 12:1 | 3.9153 | 12.5230 |

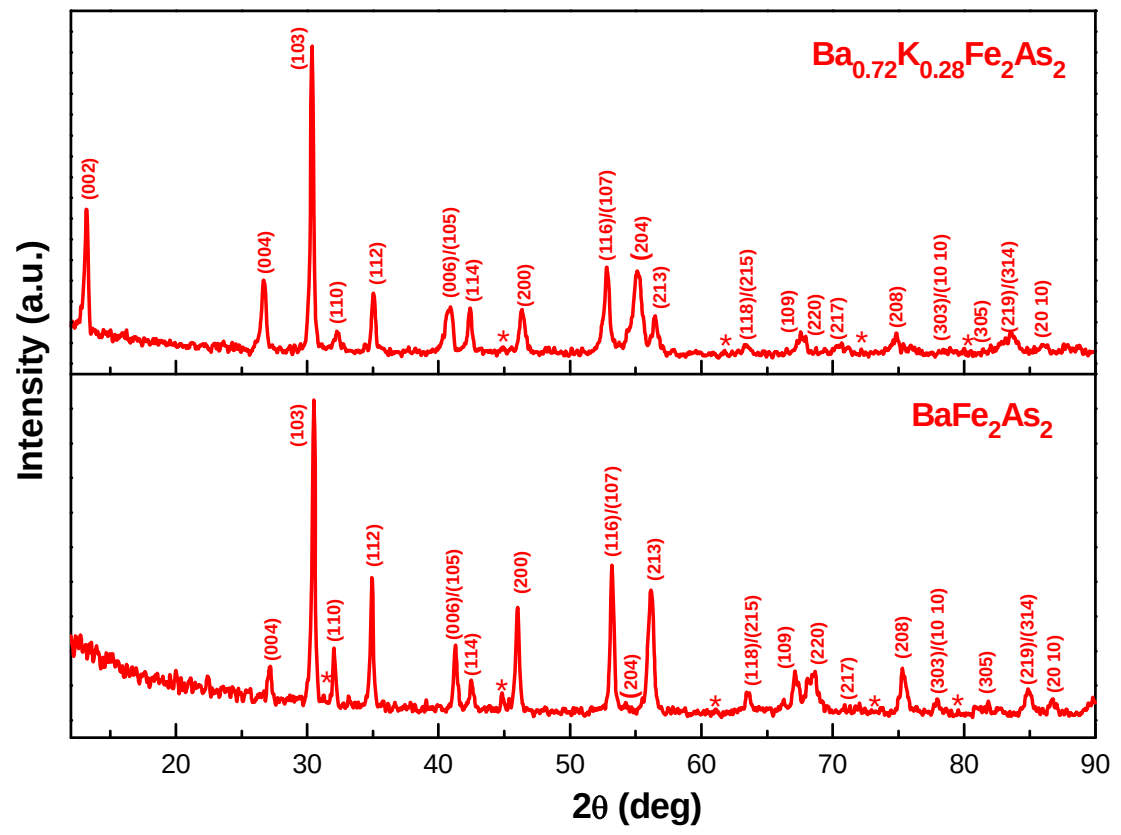

(a)

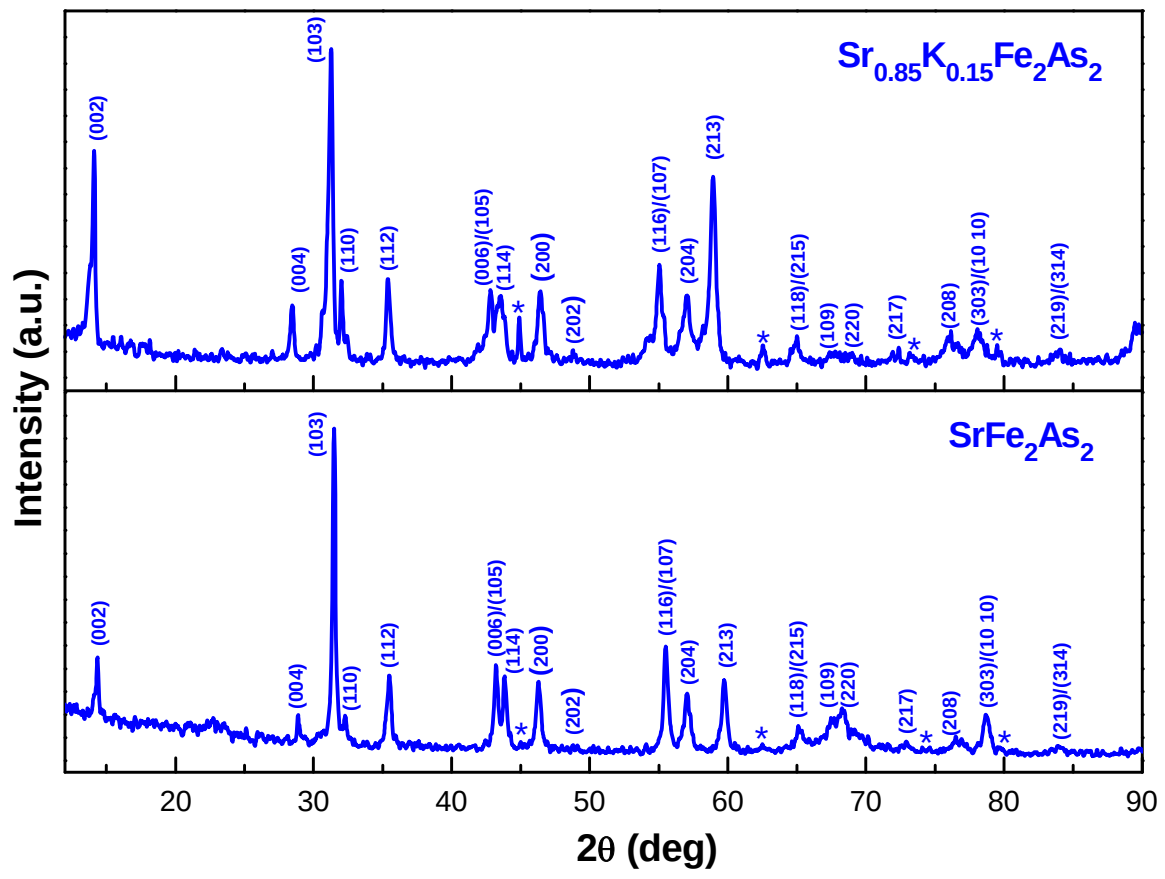

(b)

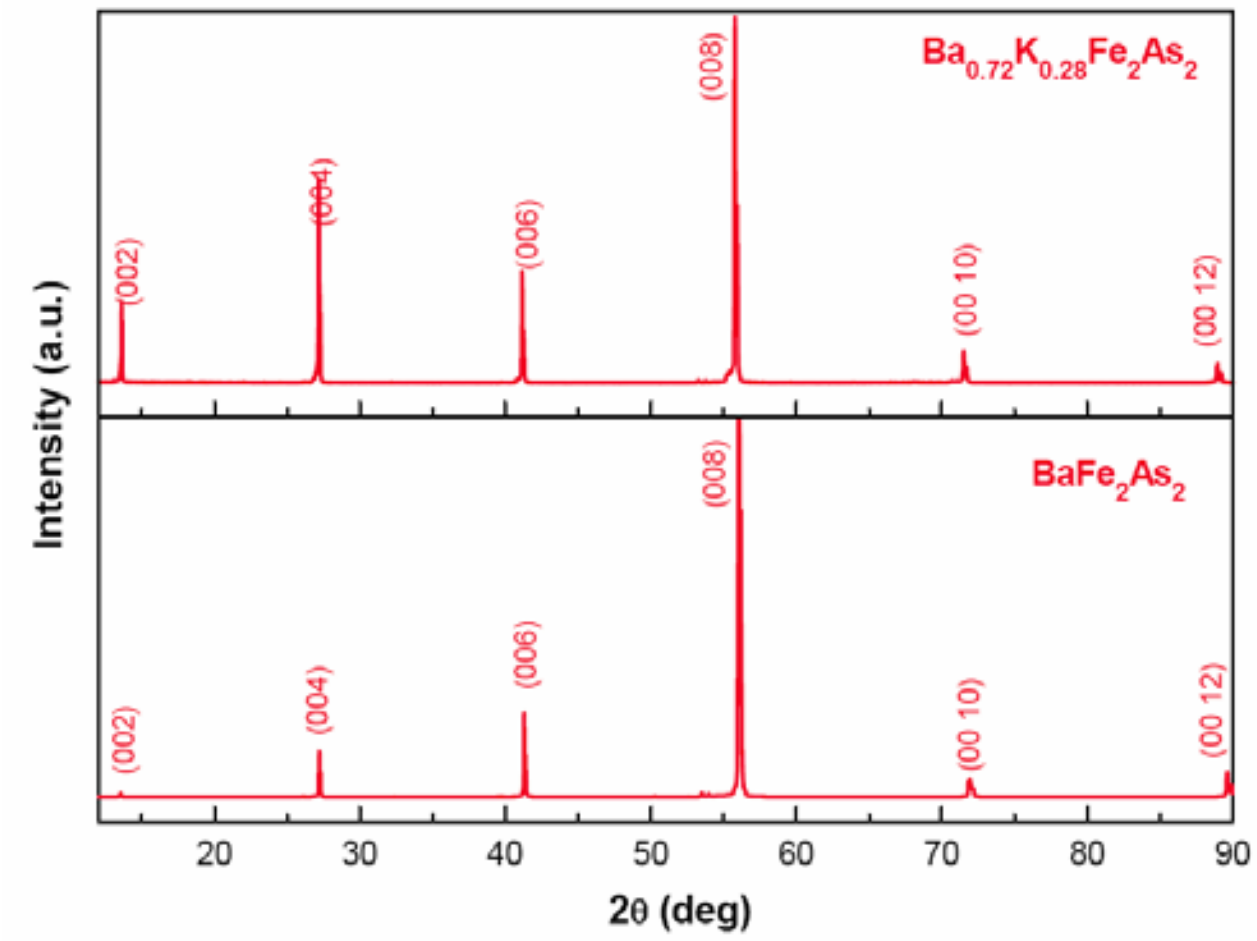

(c)

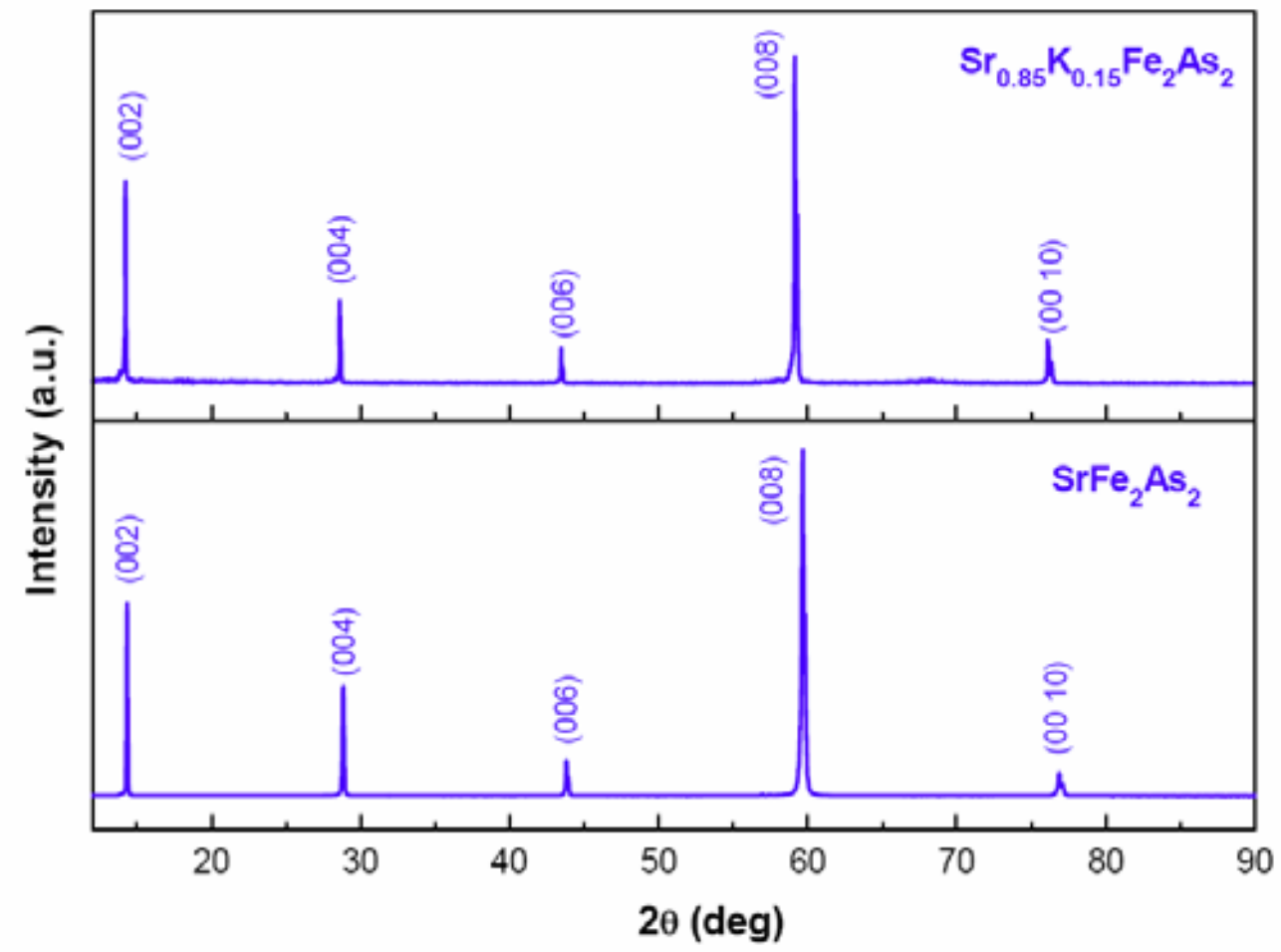

(d)

Fig. 4 The XRD diffraction patterns for $A_{1-x}K_xFe_2As_2$ (A=Ba, Sr) single crystals. (a) Data from pure and doped $Ba_{1-x}K_xFe_2As_2$ single crystal prepared powders, (b) pure and doped $Sr_{1-x}K_xFe_2As_2$ single crystal prepared powders, (c) typical *(00l)* patterns taken on the pure and doped $Ba_{1-x}K_xFe_2As_2$ single crystals, and (d) the *(00l)* patterns taken from pure and doped $Sr_{1-x}K_xFe_2As_2$ single crystals. Peaks due to the impurity phase of Sn are marked by "*".

*3.4 Crystal habit*

Owing to the layered structure, the pnictide $A_{1-x}K_xFe_2As_2$ crystals exhibit a platelet-like morphology and the *c*-axis is normal to the (*001*) face. The crystal facets can well develop according to the crystallographic orientation. The Miller index faces of the as-grown crystals were identified using a two-circle optical goniometer. The typical crystallization facets are shown in Fig. 3(b). The crystal exhibits the tetragonal structure comprised the main index faces of (*001*), (*013*) and (*010*) around the crystallization axis of [010]. According to the developed facets, the crystals show an anisotropic growth behavior. The growth rate along the (*010*) direction is faster than the (*001*), because the atomic bonding energy is higher for the *a* than the *c* axis. Along the *a* direction the total number of bonds is 16, i.e., Ba-As (8), Fe-As (4) and Fe-Fe (4), while along the *c* axis there is only Ba-As (8), for the case of $BaFe_2As_2$ [19]. Doping of $K^+$ (1.51 Å) causes a decrease of As-Fe-As bond angle and an increase of the bond length, leading to a weakening the bonding energy. This can result in a lower growth rate of the *c* direction and forming even thinner platelet of the crystals.

There is an interesting crystal habit for all such pnictide compounds that exhibit multilayer stacks, macrosteps, step bunches and inclusions in the (*001*) face. These typical features are shown in Figs. 5(a)-(f). Fig. 6(a) shows an edge of crystal exposing multiple layers for each thickness of ~5μ. These crystal layers are readily cleaved mechanically. Fig. 5(b) is a winded sheet of the cleaved layer due to its metallic-like behavior. Macro-steps are observed on the (*001*) plane, as shown in Fig. 5(c). The step terraces were formed at an uncompleted growth state. The frontier of the steps has a tetragonal structure with the well developed facets of (100)/(010), as shown in Fig. 5(d). These macrosteps spread laterally along the [100]/[010] direction, and then a new growth center occurred  on the grown layer of (*001*). No growth spirals were observed in the (*001*) faces. It is indicated that the crystal growth on the (*001*) face occurs at a relatively high supersaturation according to two-dimensional nucleation, layer-by-layer growth mechanism.

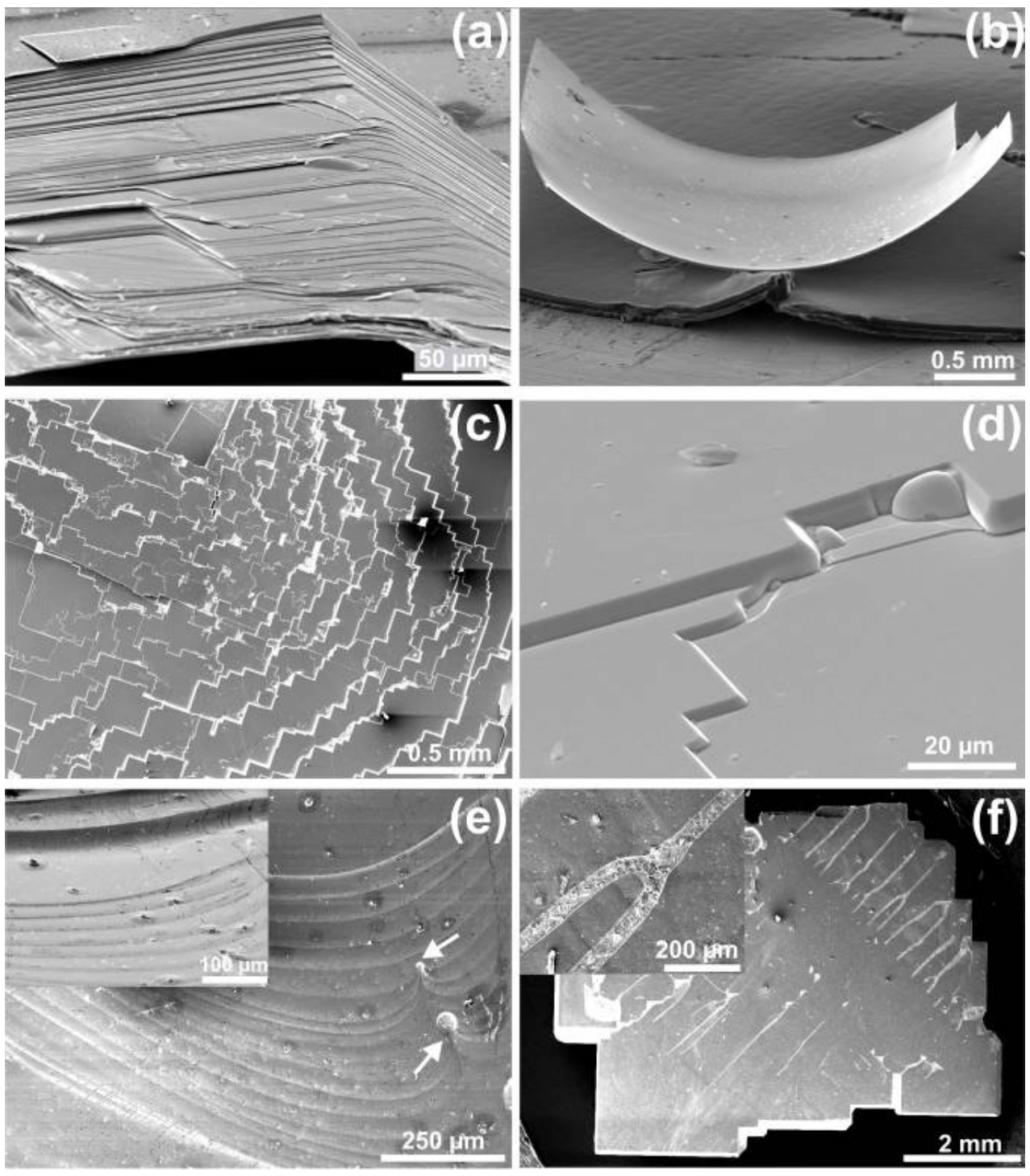

Fig. 5: SEM micrographs of (a) $SrFe_2As_2$ with multilayer stacks, (b) a winded sheet of $Ba_{0.72}K_{0.28}Fe_2As_2$ single crystal, (c) macrosteps on the (*001*) face of $Ba_{0.72}K_{0.28}Fe_2As_2$, (d) and the enlarged part of (c), (e) the step bunching of $SrFe_2As_2$, the arrows show the retardation of steps by the Sn impurity particles, (f) the inclusions of Sn formed in the (*001*) of $SrFe_2As_2$ crystal. Inset is an enlarged part. The crystals were grown from Sn-flux.

Precipitates of Sn give rise to step bunches on the (*001*) face. These bunches can cause curved shapes at the crystal growing fronts and finally stop developing crystal facets. The typical punches are indicated by arrows in Fig. 5(e). The Sn precipitates could be trapped as inclusions and lead to form secondary phases in crystals. Fig. 5(f) shows the Sn inclusion lines as impurities formed in the (*001*) face of the $SrFe_2As_2$ crystal grown from the melt in a supercooled state. The content of Sn impurities in the crystals are determined to be <1.0 at% by both EDX and ICP.

### *3.5 Transport and magnetic properties*

#### *3.5.1 Different features in $BaFe_2As_2$ crystals grown by Sn and self-flux*

The temperature dependence of the in-plane resistivity for $BaFe_2As_2$ single crystals was measured between 4 and 300 K under zero magnetic field. The results are presented in Fig. 6(a). A weak temperature dependence is seen from 138 to 300 and 85 to 300 K, indicating metallic behavior at higher temperatures for both $BaFe_2As_2$ crystals formed in self- and Sn solvent, respectively. A sharp decrease is seen at 138 K for the crystals grown by self-flux, while it occurs at 85 K for the crystals obtained by Sn flux. Such behaviour is similar to the transition in LaFeAsO driven by the spin density (SDW) instability in the iron layers [20, 21]. A pronounced difference in the spin density wave anomaly is observed comparing the two crystals, which might be attributed to the tiny amount of Sn incorporated into the crystals and related to a reduction in $T_s$. As cooling down below 85 K the resistivity increases rapidly to 50 K and then decreases again to 4 K. The resistivity is approximately 30% (at 50 K) and 20% (at 4 K) higher than it is at 85 K. A sharp superconducting transition temperature $T_c$ occurs at 38.5 K for the doped $Ba_{0.68}K_{0.32}Fe_2As_2$, while above $T_c$ there exist a nearly linear temperature dependence and a featureless increase in resistivity for higher temperatures up to room temperature. The in-plane resistivity is smaller for doped than for undoped single crystals and a sharp superconducting temperature transition occurs at $T_c$~38.5 K, as shown in Fig. 6(b). The spin density wave is entirely suppressed by doping [17].

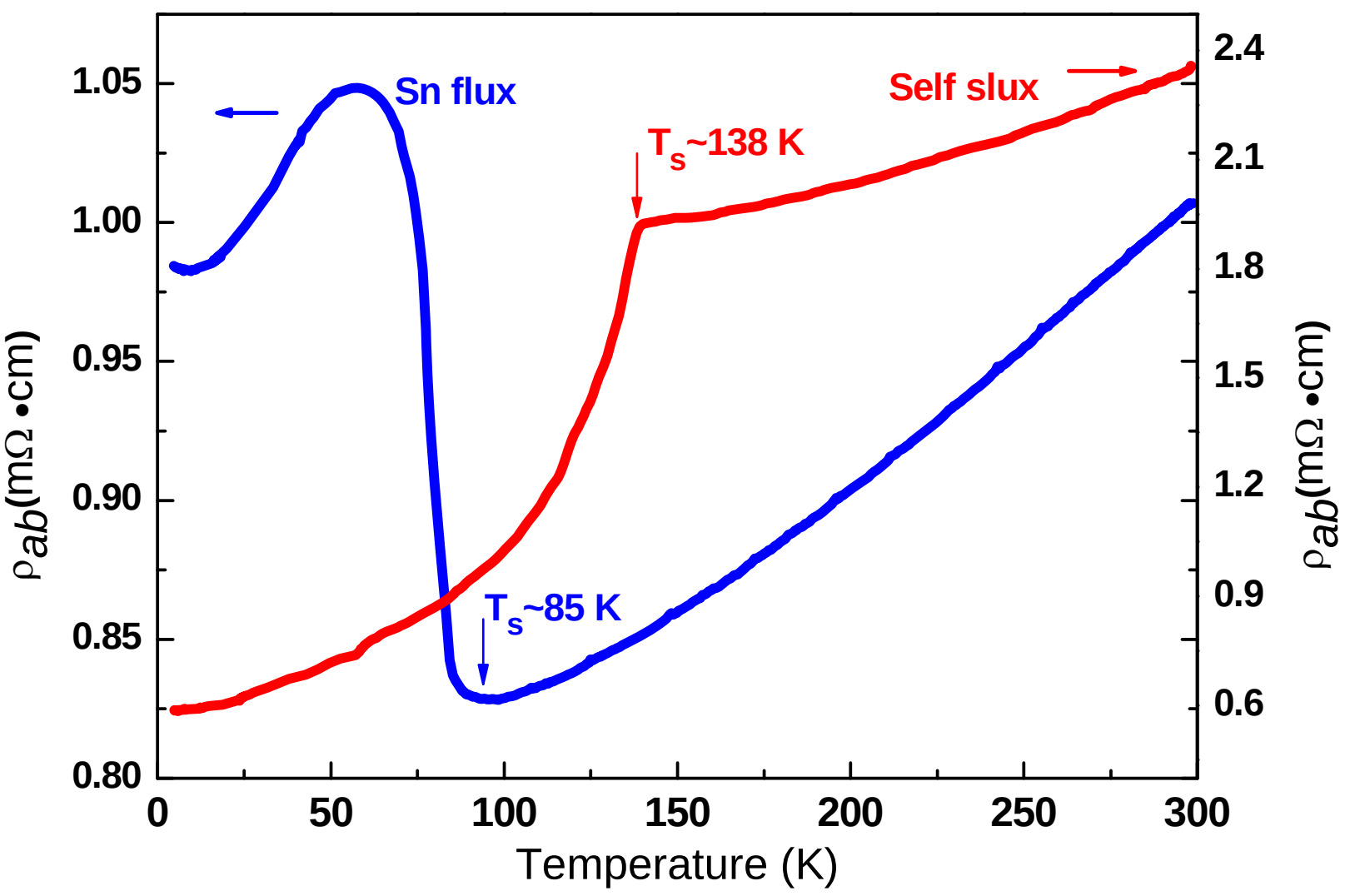

(a)

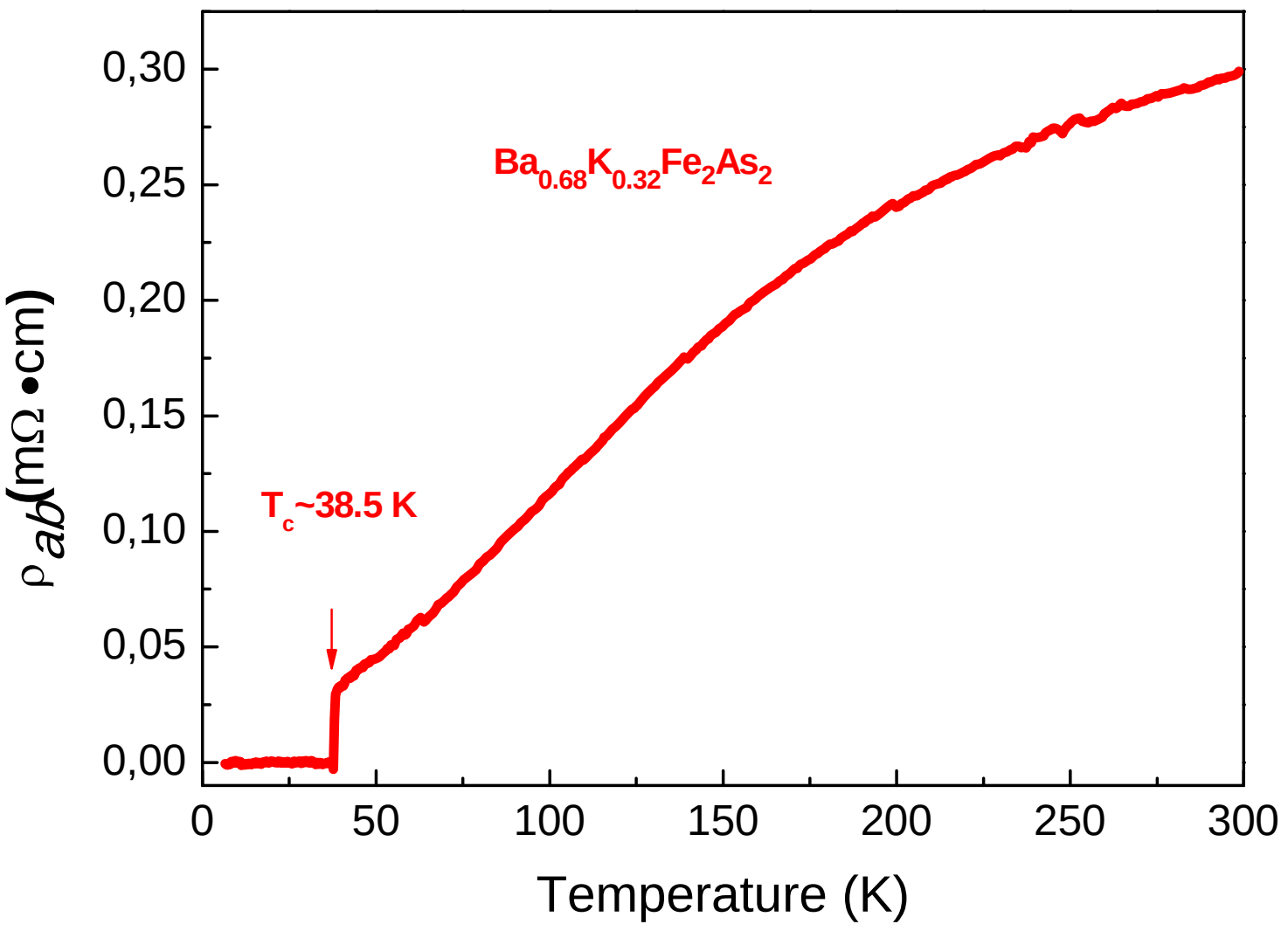

(b)

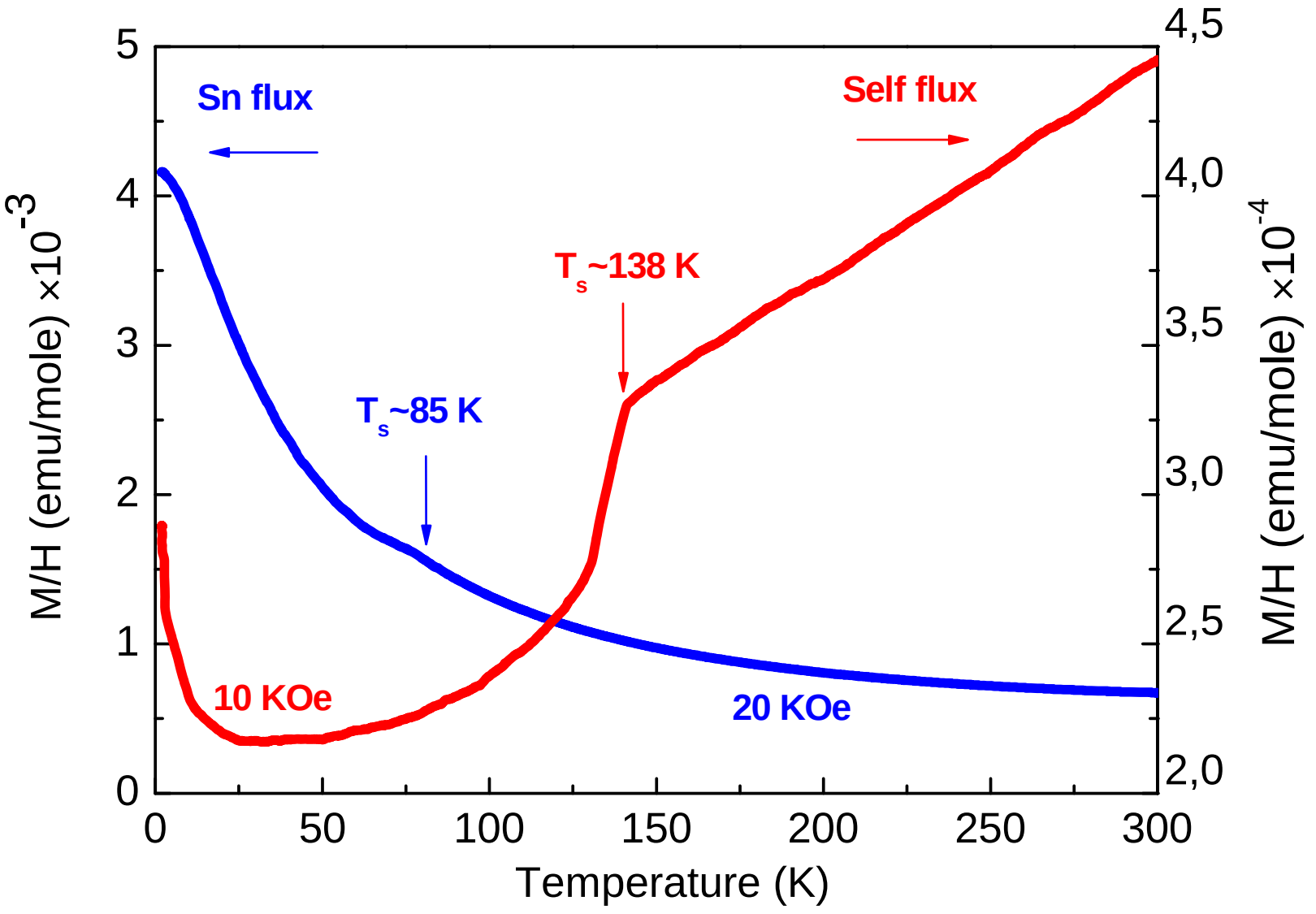

(c)

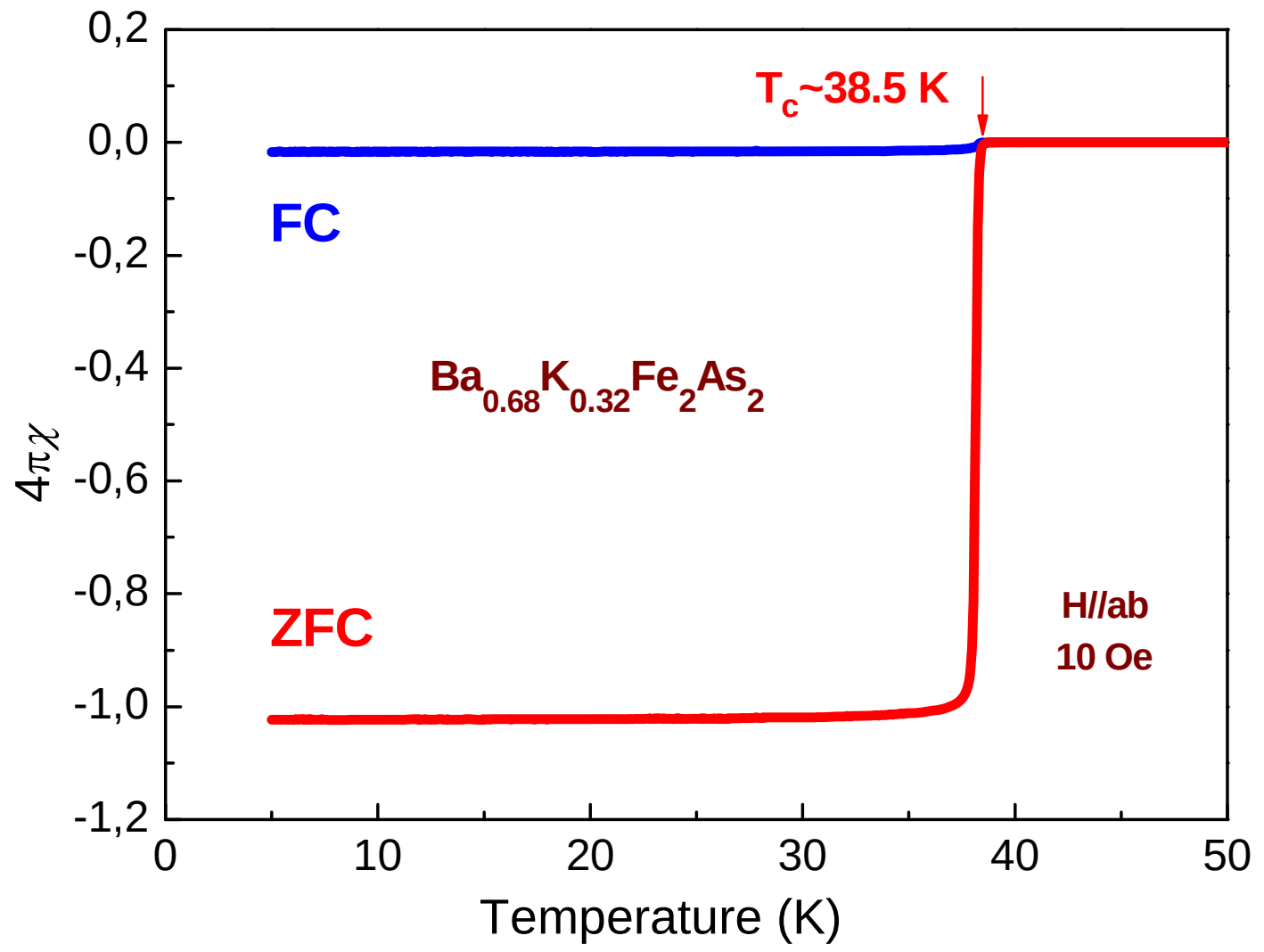

(d)

Fig. 6 (a) Temperature dependence of the in-plane resistivity for $BaFe_2As_2$ single crystals formed by self- and Sn flux, respectively. Different features are observed at $T_s$ in crystals obtained using different flux. (b) Temperature dependence of the in-plane resistivity for the doped $Ba_{0.68}K_{0.32}Fe_2As_2$ single crystal obtained by self-flux. (c) Temperature dependence of magnetization data with the field H//*c* for the $BaFe_2As_2$ and doped (self-flux) $Ba_{0.68}K_{0.32}Fe_2As_2$ single crystals demonstrate a superconducting transition temperature $T_c$~38.5 K.

The magnetic susceptibility data in Fig. 6(c) show a clear transition at 85 K in the H || *c* curve. The cross point at T~85 K can be resolved by two fits by a Curie-Weiss formula to the high and low temperature parts. This behaviour is seen for the crystals grown using Sn flux but not self-flux or polycrystalline samples, which exhibit a spin density wave (SDW) transition at ~140 K [8, 9, 16, 17]. This pronounced discrepancy may be an effect associated with the basal plane of FeAs with Fe sites partly replaced by Sn. The tiny amount of Sn is likely to be incorporated on to Fe sites during the crystal formation in Sn solvent, since the ionic radius of $Fe^{2+}$ (0.78 Å) is closer to $Sn^{2+}$ (0.93 Å) instead of $Ba^{2+}$ (1.42 Å). A sharp superconducting transition measured by magnetic susceptibility for the $Ba_{0.68}K_{0.32}Fe_2As_2$ crystal is found at $T_c$~38.5 K, as shown in Fig. 6(d). The shielding fraction close to 1 demonstrates the bulk nature of superconductivity. The onset transition temperature of $T_c$~38.5 K defined by a 10% decrease from the normal

state magnetic susceptibility is the highest one ever reported. The superconducting transition width $\Delta T_c = T_c(10\%) - T_c(90\%) = 0.3$ K is the smallest ever reported.

*3.5.2 Raman scattering*

In the upper panel of Fig. 7(a) we compare Raman spectra of self-flux and Sn flux single crystals at room temperature. According to the factor group analysis of the I4/mmm space group, two Raman-active $A_{1g}$ and $B_{1g}$ modes are expected in the in-plane scattering configuration. We observe a single, sharp peak at 210 $cm^{-1}$. For both samples the linewith of this mode is small, which testifies the high quality of the single crystals. At low temperatures, a weaker peak at 183 $cm^{-1}$ is observed. The evolution of the latter modes might be due to an enhanced Raman matrix element of the respective phonon caused by the structural phase transition to the orthorhombic (*Fmmm*) phase. The maxima at 210 $cm^{-1}$ and 183 $cm^{-1}$ are assigned to $B_{1g}$ and $A_{1g}$ symmetry, respectively [22]. The $B_{1g}$ mode corresponds to a displacement of Fe atoms along the c axis and is thus susceptible to any change of the Fe-*d* states around the Fermi level.

With decreasing temperatures the 183 $cm^{-1}$ mode of the Sn flux crystal gains intensity while the 210 $cm^{-1}$ mode undergoes a shift in frequency and displays a narrowing in linewidth, see Fig. 7(a). The Raman spectra of the self-flux single crystals are very similar (not shown here). A fit to a Lorentzian profile leads to frequency and linewidth that are compared in Fig. 7(b) for both crystals. Upon cooling from room temperature to $T_s$ ~138 K, the phonon mode hardens by 3 $cm^{-1}$ for both crystals. For the self-flux single crystal the 210 $cm^{-1}$ mode shows a discontinuous jump by 1 $cm^{-1}$ around $T_s$ and then undergoes a hardening by 1 $cm^{-1}$ between $T_s$ and 4 K. Its linewidth shows a substantial change around $T_s$ as well; it drops exponentially with decreasing temperature below $T_s$. The simultaneous, discontinuous change of the phonon frequency and linewidth corroborates a first-order nature of the structural phase transition for the self-flux single crystal.

As to the Sn flux sample, the temperature dependence of the 210 $cm^{-1}$ mode agrees with that of the self-flux sample in the temperature range between room temperature and $T_s$ except a broader linewidth. This is associated with the shortening of a phonon lifetime due to Sn impurities. Below $T_s$ ~138 K marked differences are observed. The onset temperature of the phonon anomalies shifts down to 85 K, consistent with the transport and magnetic measurements shown above. Around 85 K the jump-like feature of the phonon frequency disappears whereas only the small change of the slope is seen. For temperature below 85 K the exponential decrease of the linewidth is rounded out and its magnitude is three times smaller than that of the self-flux sample. This suggests that the presence of the Sn impurities suppresses the first-order transition and drives the system to a transition of second-order.

In the following we estimate the anharmonic phonon contribution to the phonon frequency using a phonon-phonon interaction mechanism [23]

$$\omega_{ph}(T) = \omega_0 + C\left(1 + \frac{2}{e^{\hbar\omega_0/2k_B T} - 1}\right),$$

where $\omega_0$ is the frequency of the $B_{1g}$ mode at zero temperature and C is a constant, related to the decay of an optical phonon into acoustic phonons. The fit to the experimental data with $\omega_0$=217 cm$^{-1}$ and C=-1.8 cm$^{-1}$ is given by the dashed line. The phonon-phonon mechanism provides a reasonable description in the temperature range of 138 – 295 K. However, there exist noticeable discrepancies for temperatures below $T_s$. This necessitates an additional mechanism. In a SDW system electron-phonon coupling can produce substantial phonon anomalies. Here we note that even below $T_s$ the difference between the fitted and experimental data is not large. This implies that electron-phonon coupling is not pronounced for the $A_{1-x}K_xFe_2As_2$ family. Nonetheless, due to this coupling the phonon linewidth is extremely susceptible to any change in the electronic density of states at the Fermi level and thus can serve as a measure of the free carrier concentration [21]. In our case, the temperature dependence of the linewidth resembles that of the resistivity at least for the self-flux sample [compare Fig. 6(a) and Fig. 7(b)]. The exponential drop of the linewidth below $T_s$ is therefore taken as a characteristic of the opening of the SDW gap on the Fermi surface. The drop is still observed for the Sn flux sample while the transport experiment is dominated by Sn impurities. This implies that the SDW gap is still open on the Fermi surface of the Sn flux sample. However, structural distortions are weaker or incoherent and details of the magnetic and electronic structure might be different from the "undoped" system.

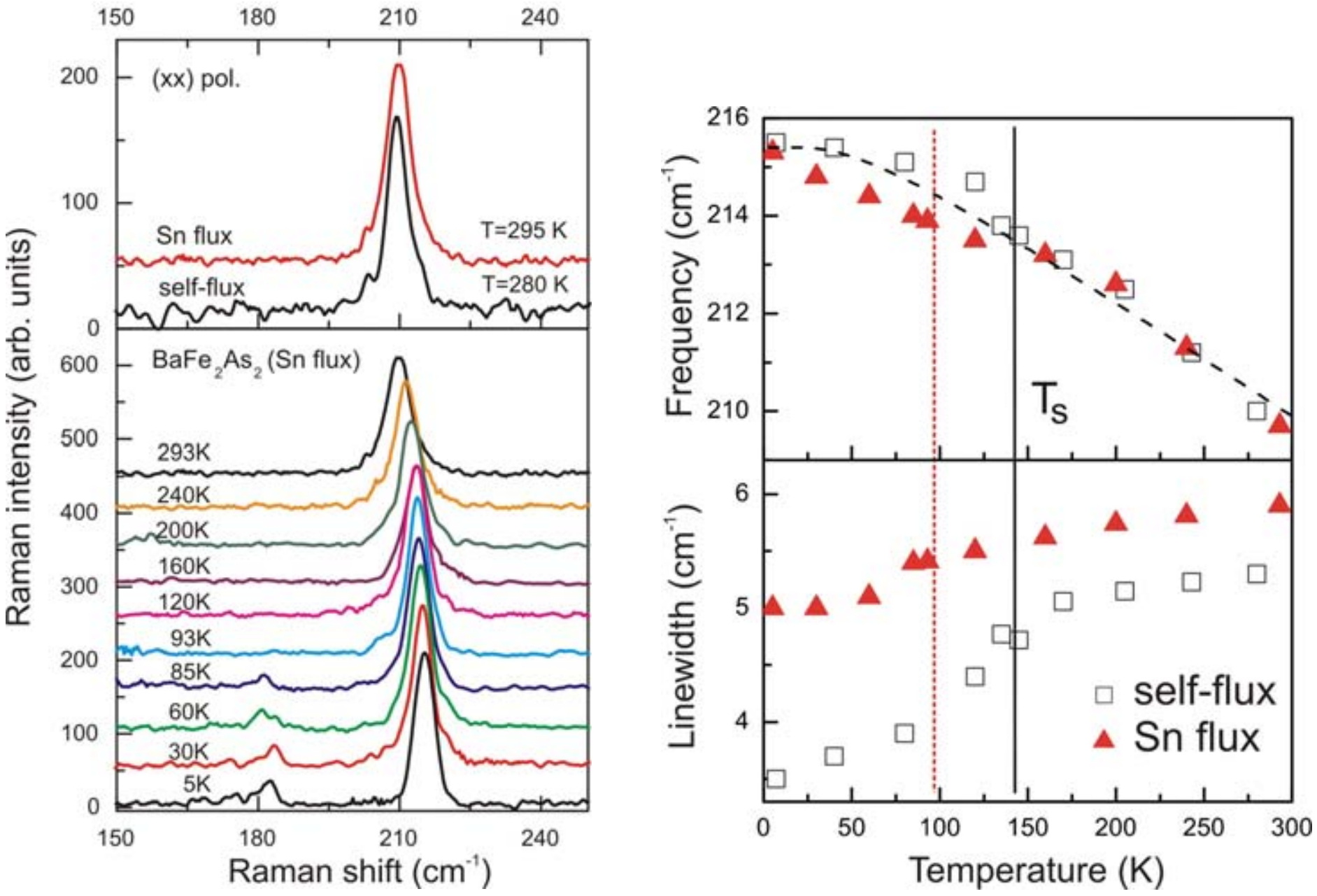

(a) (b)

Fig. 7 (a) (Upper panel) Comparison of Raman spectra of the self-flux and Sn flux samples at room temperature. (Lower panel) Temperature dependence of Raman spectra of the Sn flux sample. (b) Temperature dependence of the peak position and linewidth of the 210 $cm^{-1}$ mode for both samples. The dashed line in the upper graph is a fit to the experimental data using a phonon-phonon interaction model (see the text for details).

### *3.5.3 Features of pure and doped $SrFe_2As_2$*

The in-plane resistivity of $SrFe_2As_2$ shows a linear temperature dependence for temperatures above 202 K with a metallic behaviour, as shown in Fig. 8 (a). The observed kink is attributed to the onset of the spin density wave (SDW) transition associated with a structural transition from tetragonal to orthorhombic at ~200 K [8, 9, 16, 17, 20, 21]. It is noticed that no change of the SDW $T_s$ is observed in the crystals formed by Sn flux, compared to that formed in self-flux, although tiny Sn contaminations are seen in Fig. 4(b). For the potassium doped $Sr_{0.85}K_{0.15}Fe_2As_2$ there is an anomaly occurring at 178 K for the SDW and at 32 K for the superconducting transition. This suggests a coexistence of superconductivity and SDW in the weakly doped regime, which is consistent with angle resolved photoemission spectroscopy data [24]. Hole- or electron-doping can suppress the temperature anomaly and introduce superconductivity [16, 17, 25-27]. Our sample of $Sr_{0.85}K_{0.15}Fe_2As_2$ shows a sharp superconducting transition temperature $T_c$=32 K together with $T_s$=178 K, as shown in Figs. 8(a). This indicates that a K doping of x=0.15 can not completely suppress the resistance anomaly. Thus superconductivity can coexist with the orthorhombic structure and the antiferomagnetic state below $T_c$, which was also proven in $Ba_{1-x}K_xFe_2As_2$ [16].

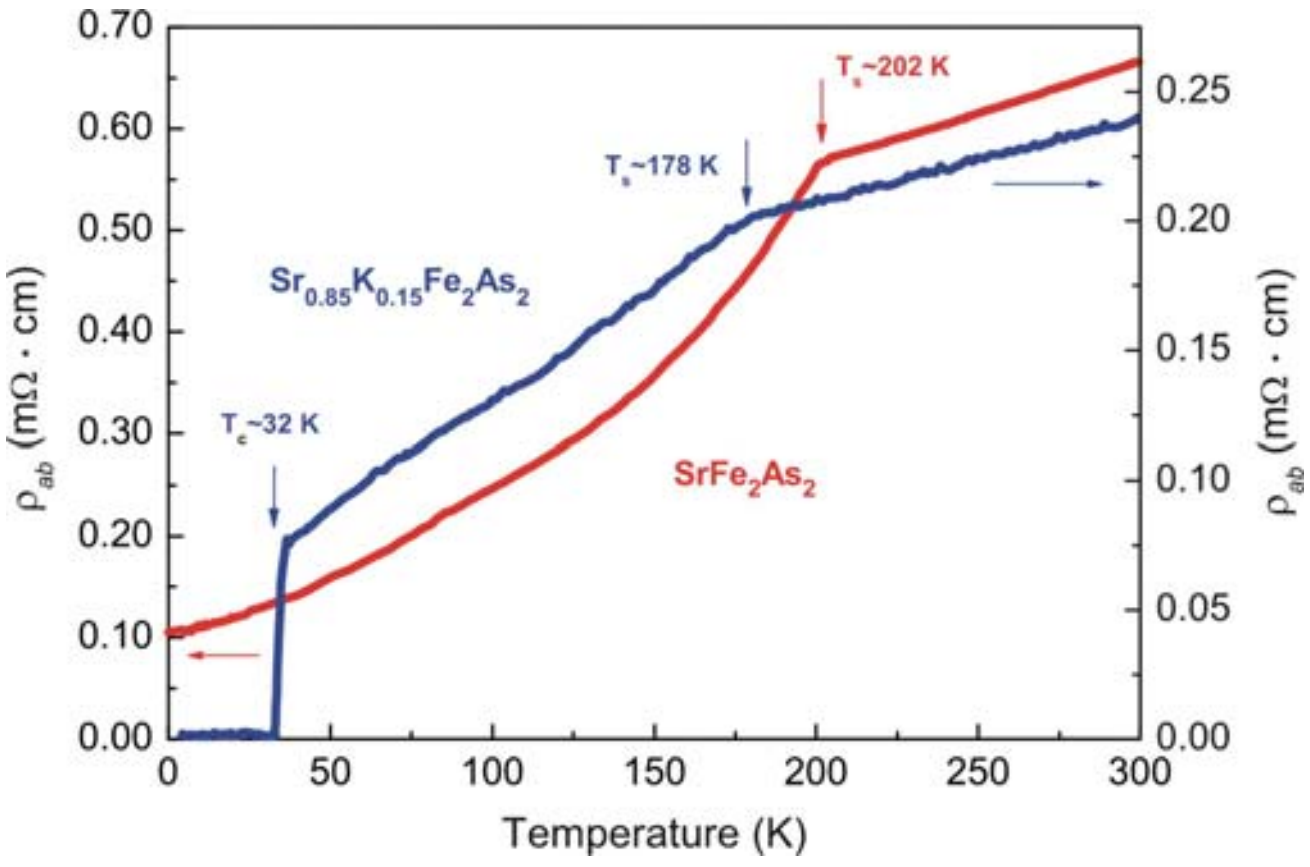

(a)

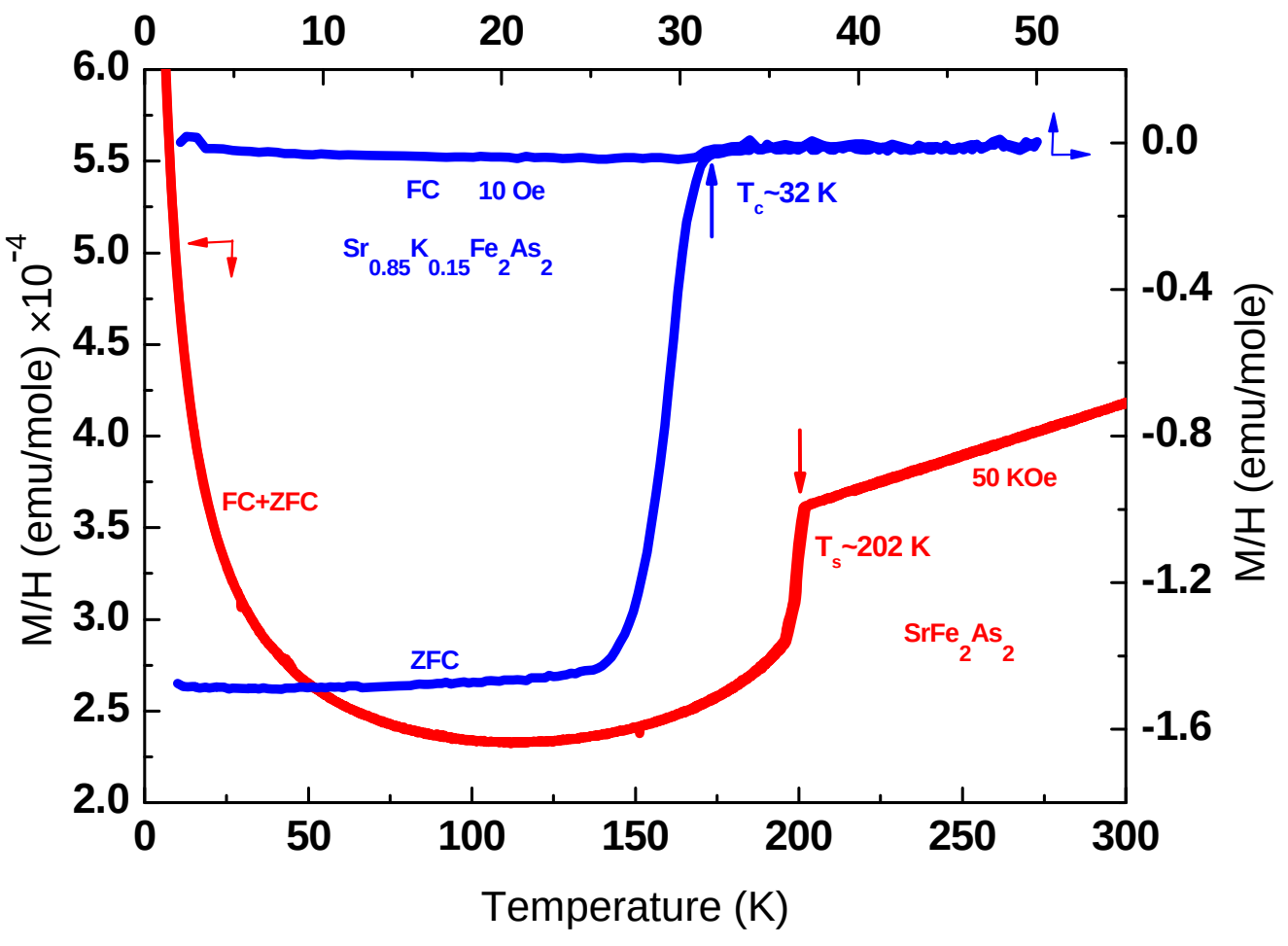

(b)

Fig. 8 (a) Temperature dependence of the in-plane resistivity for the pure $SrFe_2As_2$ with an anomaly at $T_s$~202 K and a coexistence of superconductivity with $T_c$~32 K and spin density wave with $T_s$~178 K for the weakly doped $Sr_{0.85}K_{0.15}Fe_2As_2$ single crystal. (b) Temperature dependence of magnetization data with the field H//*c* for $SrFe_2As_2$ and $Sr_{0.85}K_{0.15}Fe_2As_2$. The crystals were grown by Sn-flux.

The magnetic data in Fig. 8(b) show a further confirmation of the superconducting transition temperature $T_c$~32 K for the $Sr_{0.85}K_{0.15}Fe_2As_2$ single crystal. For the pure $SrFe_2As_2$, the magnetic susceptibility decreases as the sample is cooled with a small slope to $T_s$~202 K. A minimum temperature range between 100 and 150 K is observed and assistant to the samples obtained by Sn flux [9]. Below ~100 K the susceptibility becomes more anisotropic with a clear Curis-Weiss-like tail, which is not affected by Sn. Nerveless the doping of K supresses these features.

*3.5.4 High resolution photoelectron spectra*

It is known that Sn is included into the crystal structure and leads to a modification of the SDW state. However it is not known which site Sn occupies in the structure. We applied ESCA spectroscopy recorded with an electron spectrometer (AXIS ULTRA, Kratos, UK) with monochromatized $AlK_\alpha$ radiation. The vacuum was kept at below $1x10^{-9}$ mbar during the measurements. Photoelectron peaks of As 2*p*,3*s*, 3*p*, and 3*d*, Ba 3*p*, 3*d*, 4*p*, and 4*d*, Fe 2*p*, O 1*s*, Sn 3*d*, C 1*s*, Sr 3*p*, were recorded. In addition, K 2*p* peaks were measured for K doped sample. The surface of all samples was cleaned by sputtering 1 kV $Ar^+$ ions for 1 minute.

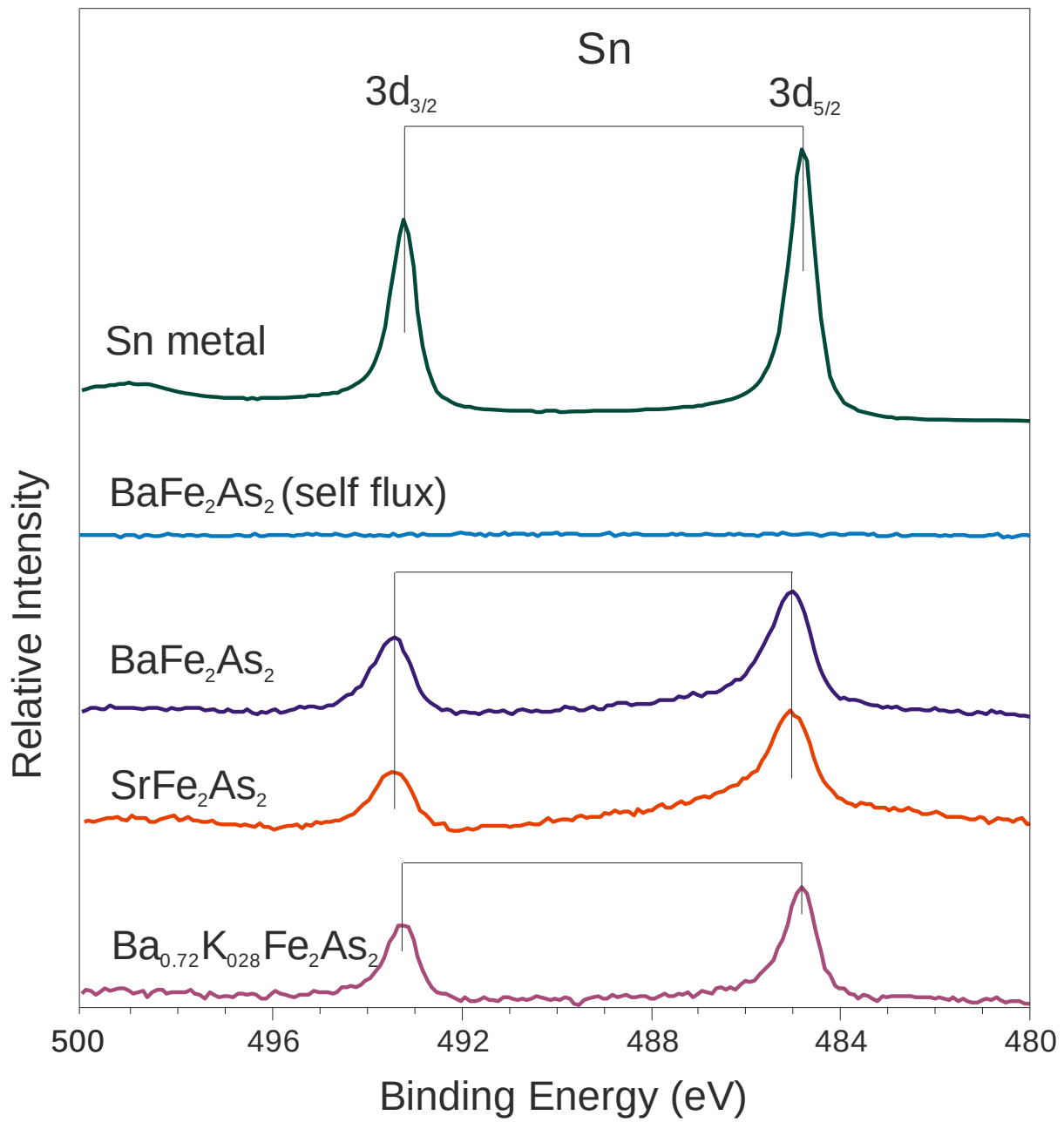

Fig. 9 photoelectron spectra for Sn, $BaFe_2As_2$, $SrFe_2As_2$, and $Ba_{0.72}K_{0.28}Fe_2As_2$ single crystals.

Fig. 9 shows Sn 3*d* photoelectron spectra for $BaFe_2As_2$, $SrFe_2As_2$, and $Ba_{0.72}K_{0.28}Fe_2As_2$ grown by Sn flux, as well as $BaFe_2As_2$ grown by self flux method. All measured crystals formed by Sn flux contain ~1 at% of Sn. The Sn 3*d* photoelectron spectrum of Sn metal is also shown in order to compare the peak positions (chemical shift in binding energy). For metallic Sn, the $3d_{5/2}$ peak was observed at the binding energy of 484.8 eV, which is in good agreement with the values reported in literature. No Sn was detected in $BaFe_2As_2$ grown by the self flux method. Sn $3d_{5/2}$ photoelectron peaks were observed at 485.1 and 485.0 for BaFeAs and SrFeAs crystals, and 484.9 eV for K-doped $Ba_{0.72}K_{0.28}Fe_2As_2$. Small positive chemical shifts, +0.1~+0.3 eV suggest that Sn occupies the positive ion $Fe^{2+}$ sites in the basal plane of the crystals.

## 4. Conclusion

In conclusion, large $A_{1-x}K_xFe_2As_2$ (A=Ba, Sr) single crystals with sizes in the centimeter range have been grown using either from Sn or self flux in zirconia crucibles. By a specially designed furnace inset, as-grown crystals can be readily separated from the residual flux. Well developed crystallographic facets of (*001*), (*013*) and (*010*)/(*001*) are observed on as-grown crystals. The morphology of the crystals exhibits a growth anisotropy with a much faster growth rate along the [010] than the [001], resulting in a layer structure. The discrepant spin density wave anomaly at $T_s$ ~138 and 85 K that occurs in $BaFe_2As_2$ crystals grown using self and Sn flux, respectively, is based on the substitution of the Fe within the FeAs basal plane. The presence of Sn impurities not only suppress the ordering temperature but also transforms the first-order transition to a second-order one. We find that the electronic and magnetic properties of the $A_{1-x}K_xFe_2As_2$ family are extremely sensitive to the extrinsic influence of the Sn impurities. This restricts the universal role of the SDW in explaining superconductivity. Doping of K supresses the spin density wave and induces superconductivity in the doped $A_{1-x}K_xFe_2As_2$. We evidence a co-existence of SDW and superconductivity in the lightly doped $Sr_{0.85}K_{0.15}Fe_2As_2$.

**Acknowledgement**

The authors thank D.C. Johnston for important discussions, G. Götz and Z. Ioannis for X-ray diffraction measurements, and E. Brücher for the SQUID characterization, C. Bush for the analyses of crystal composition, and L. Dorner-Finkbeiner and H. Bender for technical support.